\documentclass[journal]{IEEEtran}
\IEEEoverridecommandlockouts
\usepackage{cite}
\usepackage{amsmath,amssymb,amsfonts}
\usepackage{algorithmic}
\usepackage{graphicx}
\usepackage{textcomp}
\usepackage{xcolor}
\usepackage{caption}
\usepackage{subfigure}
\usepackage{booktabs} 
\usepackage{verbatim}
\usepackage{amsmath}
\usepackage{mathtools}
\usepackage{multirow}
\usepackage[linesnumbered,boxed,ruled,commentsnumbered]{algorithm2e}
\def\BibTeX{{\rm B\kern-.05em{\sc i\kern-.025em b}\kern-.08em
    T\kern-.1667em\lower.7ex\hbox{E}\kern-.125emX}}
\usepackage{authblk}
\usepackage{bbding}
\usepackage{footnote}
\usepackage{threeparttable} 
\usepackage{makecell}
\usepackage{color,soul}

\begin{document}

\title{Deep Reinforcement Learning-Based Bidding Strategies for Prosumers Trading in Double Auction-Based Transactive Energy Market}

\author{Jun Jiang, Yuanliang Li, Luyang Hou, Mohsen Ghafouri, Peng Zhang, Jun Yan, Yuhong Liu
        % <-this % stops a space
\thanks{J. Jiang and Y. Liu are with the Department of Computer Science and Engineering, Santa Clara University, Santa Clara, CA, USA (e-mail: jun3525114@gmail.com, yhliu@scu.edu) }
\thanks{Y. Li, M. Ghafouri, and J. Yan are with Concordia Institute for Information Systems Engineering, Concordia University, Montreal, QC, Canada (e-mail: \{yuanliang.li, mohsen.ghafouri, jun.yan\}@concordia.ca) }
\thanks{L. Hou is with Beijing University of Posts and Telecommunications, Beijing, China (e-mail: luyang.hou@bupt.edu.cn) }
\thanks{P. Zhang is with the College of Information Engineering, Shenzhen University, Shenzhen, China (e-mail: zhangp@szu.edu.cn) }
}

\maketitle

\begin{abstract}
With the large number of prosumers deploying distributed energy resources (DERs), integrating these prosumers into a transactive energy market (TEM) is a trend for the future smart grid. A community-based double auction market is considered a promising TEM that can encourage prosumers to participate and maximize social welfare. However, the traditional TEM is challenging to model explicitly due to the random bidding behavior of prosumers and uncertainties caused by the energy operation of DERs. Furthermore, although reinforcement learning algorithms provide a model-free solution to optimize prosumers' bidding strategies, their use in TEM is still challenging due to their scalability, stability, and privacy protection limitations. To address the above challenges, in this study, we design a double auction-based TEM with multiple DERs-equipped prosumers to transparently and efficiently manage energy transactions. We also propose a deep reinforcement learning (DRL) model with distributed learning and execution to ensure the scalability and privacy of the market environment. Additionally, the design of two bidding actions (i.e., bidding price and quantity) optimizes the bidding strategies for prosumers. Simulation results show that (1) the designed TEM and DRL model are robust; (2) the proposed DRL model effectively balances the energy payment and comfort satisfaction for prosumers and outperforms the state-of-the-art methods in optimizing the bidding strategies.
\end{abstract}

\vspace{2mm}

\section{Introduction}
\IEEEPARstart{W}{ith} the extensive deployment of energy storage systems, solar photovoltaics (PVs), smart home appliances, and information technology, passive consumers in the traditional electricity market are gradually converted to active prosumers (producers + consumers) with distributed energy resources (DERs), who can monitor and control energy generation, consumption, storage, and transaction to achieve specific goals, such as balancing energy costs and user comfort levels \cite{qiu2017optimal, morstyn2018using, fairley2004unruly}. However, the bi-directional energy and information flow, as well as the variability of distributed renewable energy, raises great challenges in the operation of power systems in a flexible and economically efficient way \cite{shakoor2017roadmap}. 

Therefore, it is essential to establish an alternative market that can more effectively increase prosumers' economic benefits and reduce the distribution system's peak demand. Recently, the transactive energy market (TEM) has been introduced as an innovative energy management market to incentivize prosumers' participation and reduce stress on the distribution systems \cite{trivedi2022community, foruzan2018reinforcement}. The authors in \cite{kebriaei2009simultaneous, long2018peer, abdella2018peer, qiu2022mean} adopt the peer-to-peer TEM, where prosumers with deficit energy act as consumers to purchase energy directly from prosumers with an energy surplus. The authors of \cite{cornelusse2019community} design a community-based TEM to maximize social welfare according to marginal pricing. Specifically, the proposed market collects the preferences of community members and centrally solves resource allocation problems. The authors of \cite{mohy2019transactive} propose a two-stage TEM. In the first stage, optimal power flow is used to clear the market. In the second stage, market participants determine their bids. In addition, to encourage prosumers to participate flexibly in TEM, packetized energy (PE) technology can be used to manage the generation and consumption of energy for prosumers in a request-reply way by encapsulating energy into modulated and routable energy packets \cite{li2022pemt}. 

However, there are severe challenges in the TEM \cite{taghizadeh2022deep, ye2021scalable}. First, due to the randomness of renewable energy generation, load consumption, and electricity prices, the energy trading strategies of prosumers may be uncertain. Second, since prosumers cannot access information about others, it dramatically increases the difficulty of making optimal energy trading decisions. Third, the scalability of the TEM also needs to be considered. 

To address the above challenges, effective solutions for energy management and trading are receiving increasing attention. The existing literature on energy trading and management is separated into two main categories. The first type of centralized method collects the information of prosumers' DERs and matches their energy generation and consumption \cite{lezama2018local, ma2018real, nizami2019multiagent}. Although such centralized approaches can theoretically provide the optimal solutions, they suffer from various limitations \cite{ye2021scalable}: (1) high communication requirements to transmit diverse and complex technical parameters from prosumers, (2) high computational costs and poor scalability caused by the centralized optimization process that relies on a large number of decision variables and constraints, and (3) the risks of prosumers' privacy breaches.

The second type of method is based on a distributed approach. Each prosumer can independently generate the optimal bidding strategies and control DERs. This distributed approach significantly alleviates communication and computing requirements and partially addresses privacy concerns \cite{shamsi2015economic}. Specifically, RL models are increasingly used to help prosumers make optimal energy transaction decisions in the market \cite{farhoumandi2021review}. In \cite{rashedi2016markov}, a discrete multi-agent Q-learning is proposed to optimize generator bidding in a non-cooperative Markov game. Because tabular-based Q-learning has many limitations (e.g., high storage and computational cost, low-dimensional and discrete state and action spaces), it is gradually replaced by deep neural network-based RL (DRL). The authors of \cite{du2021approximating, liang2020agent} introduce the multi-agent deep deterministic policy gradient (DDPG) to approximate the Nash equilibrium of the bidding game among power generation companies. In \cite{qiu2021multi}, a multi-agent RL for automated peer-to-peer energy trading in a double-side auction market is proposed. In \cite{ye2021scalable}, a scalable privacy-preserving multi-agent RL model for large-scale transactive energy trading is proposed. However, among most existing studies, the action strategies of agents participating in the TEM consist of either bidding price or operation of the appliances based on the dispatched energy quantity, potentially reducing the agents' maximum reward. Additionally, Additionally, sharing learned parameters between agents raises privacy concerns \cite{ye2022multi}.

This paper aims to address the limitations of previous approaches in optimizing strategic bidding decisions for self-interested prosumers. To maximize the overall welfare of the TEM and motivate all prosumers to participate in the TEM, the uniform double auction (UDA) market is introduced to efficiently clear energy transactions in the market. Specifically, in each transaction period, the market publishes two types of clearing information: (1) Market statistical information (e.g., the mean of bidding prices of buyers/sellers, the mean of bidding quantity of buyers/sellers, etc.) and (2) Market clearing information (e.g., market clearing price and the dispatched quantity for each agent). 

More importantly, this paper proposes a concurrent RL model based on the deep deterministic policy gradient (DDPG) for each prosumer to determine the optimal energy trading decisions. Specifically, each prosumer is first modeled as an agent; to protect each agent's privacy and improve the environment's scalability, we assume that prosumers only adopt distributed learning without sharing their private learning parameters. On the other hand, each agent learns the clearing statistics released by the market, which helps to stabilize the market. Unlike other existing works, in this paper, the action strategies include bidding prices and quantities to maximize their rewards (i.e., increase income as a seller and reduce energy payment without sacrificing comfort as a buyer). Because PE technology can increase the flexibility and effectiveness of DERs, the action space of bidding quantity is discrete. Multiple and diverse DERs are deployed to make this study more practical and general. Additionally, this work provides post-market control of energy consumption based on bidding results to prevent energy waste, such as overcooling.

The rest of this paper is organized as follows. Section II introduces the system model and problem formulations. Sections III and IV discuss the market Markov game formulation and the proposed RL framework. The experiment results are given in Section V, followed by a conclusion in Section VI. 

\section{SYSTEM MODEL AND PROBLEM FORMULATION}

\begin{figure}[tb]
\centering
\includegraphics[width=3.3in]{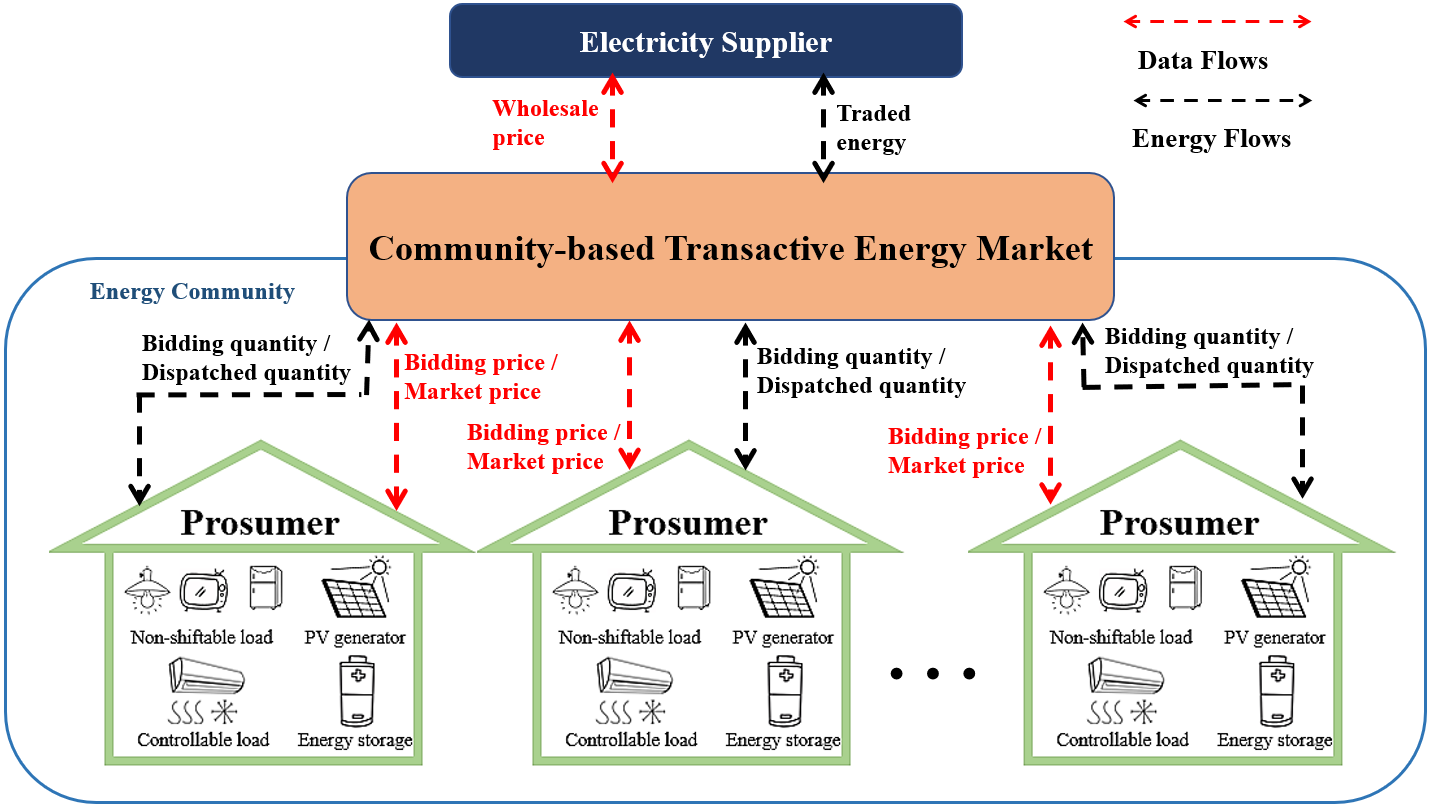}
\caption{Community-based transactive energy market overview}
\label{fig:energy_community_overview}
\vspace{-4mm}
\end{figure}

Fig. \ref{fig:energy_community_overview} depicts the transactive energy trading community, which consists of a double auction market, an electricity supplier, and numerous transactive prosumers. The double auction market is employed to manage the energy trading in the community (e.g., determining market clearing price and dispatched quantity of each participant), as well as trading with the electricity supplier. The electricity supplier trades energy to the community at its offered price. The transactive prosumers are equipped with a learning agent and an energy management system to optimize their bidding strategy (e.g., bidding price and quantity) and energy management decisions.

\subsection{Transactive Market Clearing}
Due to its good performance on individual rationality, budget balance, and economic efficiency, uniform double auction (UDA) market \cite{lin2019comparative} is adopted in this study. The auction period in the UDA market is often a fixed period. In the UDA market, there is one auctioneer and two types of traders: sellers (e.g., energy suppliers or prosumers with energy surplus) and buyers (e.g., consumers or prosumers with energy deficit). The market auctioneer is a UDA market operator whose main objective is to minimize the total payment from buyers and the total energy generation costs for sellers through a stable and uniform transaction price. 

\begin{figure}[tb]
\centering
\includegraphics[width=3.0in]{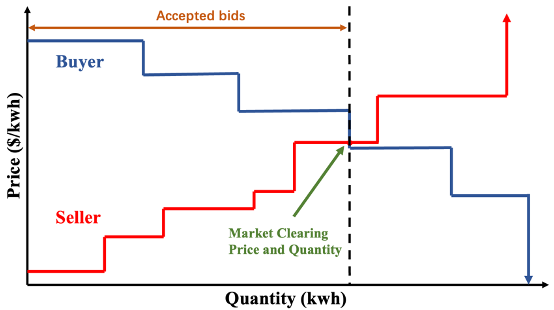}
\caption{Market clearing model}
\label{fig:Market clearing model}
\vspace{-4mm}
\end{figure}

In a UDA market, the buyers and sellers submit their bids to the market at the beginning of the auction period. Each bid contains a pair of values referring to the preferred price (i.e., $p^{b}$ for buyers or $p^{s}$ for sellers) and the amount of energy (i.e., $q^{b}$ for buyers or $q^{s}$ for sellers). After receiving bids from all traders, the market auctioneer uses the Merit-Order equilibrium model \cite{hasan2008electricity} to calculate the energy quantity supply, demand, and market transaction price. Specifically, seller bid pairs ($p^{s}$, $q^{s}$) and buyer bid pairs ($p^{b}$, $q^{b}$) are listed in ascending and descending order based on bidding prices $p^{s}$ and $p^{b}$, respectively. The intersection of the two curves determines the uniform market clearing price and quantity, as shown in Fig. \ref{fig:Market clearing model}. Then, the market auctioneer clears the market and publishes public market clearing information at the end of each auction period \cite{friedman2018double}. In the UDA market, all successful traders trade at the same clearing price. 

\subsection{Transactive Prosumer Modeling}
\subsubsection{Distributed Energy Resources Management}
In this study, the UDA-based TEM comprises a group of transactive prosumers. Prosumers reside in a feeder connected to a low-voltage substation and operate different DERs, including solar photovoltaic panels, energy storage devices, non-shiftable demand (e.g., lightning, fridge), and controllable demand (e.g., HVAC control). The DER operating parameters are driven by the preferences and requirements of prosumers and are diversified, indicating their natural variability. 

Specifically, non-shiftable demand, such as lighting and refrigerator load, is considered base load $q^{base}$. In addition, since the operation of an energy storage device (e.g., charging and discharging) significantly impacts the performance of the overall behavior of the prosumer, in this study, we assume that the prosumer house has a battery as an energy storage device. According to \cite{chen2018local}, the operation of the battery is modeled as Equation (\ref{eq:battery_opt}). The energy in battery $E^{bat}_{t}$ at step $t$ is related to the energy in battery $E^{bat}_{t - 1}$ at step $t-1$, as well as the charging power $C^{bat}_{t}$ and discharging power $D^{bat}_{t}$.
\begin{equation}
\begin{aligned}
E^{bat}_{t} = E^{bat}_{t-1} + \Delta t C^{bat}_{t-1}\alpha^{bat}_C - \Delta t D^{bat}_{t-1}/ \alpha^{bat}_D
\end{aligned}
\label{eq:battery_opt}
\end{equation}
where $\alpha^{bat}_C$ and $\alpha^{bat}_D$ represent the efficiency of battery charging and discharging respectively. The HVAC system operates by converting energy to a comfortable temperature. It can flexibly adjust the indoor temperature $T^{in}$ within the prosumer's comfortable temperature range (i.e., $T^{lb} <= T^{in} <= T^{ub}$).

\subsubsection{Cost Function of Transactive Prosumers}
This study considers two main types of costs: the costs of energy generation by solar panels and the costs of charging and discharging battery. According to \cite{ranganath2021life}, the energy cost generated by solar panels is roughly proportional to the capital cost since solar energy has no fuel costs and tiny operation and maintenance costs. Therefore, in this study, the cost of solar panels $C_{solar}$ is represented by a constant value. In addition, the energy trading decision considers the charging and discharging costs of the battery. According to the analysis of battery charging and discharging costs in \cite{das2017computationally}, the cost of the battery $C_{bat}$ is modeled as a constant value.

\subsubsection{Valuation Function of Transactive Prosumer}
The valuation function evaluates the intrinsic worth of the energy used by the HVAC system and battery for a specific prosumer $i$ at the auction round $t$. In other words, it maps energy use satisfaction to economic indicators, which implies that prosumer $i$ tend to pay for their controllable demand $q_{i,t}^{need}$. Because the base load $q^{base}$ is always satisfied, we assume that each prosumer's base load can be successfully obtained from various sources (e.g., electricity suppliers). The quantity ($q_{i,t} = q_{i,t}^{need} = q_{i,t}^{hvac} + q_{i,t}^{bat}$) in the valuation function only considers energy used by the HVAC system $q_{i,t}^{hvac}$ and battery $q_{i,t}^{bat}$. According to \cite{maharjan2013dependable, bergemann2019dynamic}, the evaluation of the prosumer $i$ is modeled as a logarithmic function of the energy used by controllable demand, as shown in Equation (\ref{eq:buyer valuation}).

\begin{equation}
\begin{aligned}
v_{i, t}(q_{i,t}) = \log(1 + \beta_i \cdot q_{i,t} (\frac{r^{bat}_{i, t}}{SOC^{bat}_{i, t}} + \frac{r^{hvac}_{i, t}}{T^{comf}_{i, t}}))
\end{aligned}
\label{eq:buyer valuation}
\end{equation}
where $r^{bat}_{i, t}$, $r^{hvac}_{i, t}$ represent the ratio of bidding quantity $q_{i,t}$ for battery and HVAC system respectively, $\beta_i$ is a positive constant value, $SOC^{bat}_{i, t}$ is the state of charge (SOC) of the battery, and $T^{comf}_{i, t}$ measures the temperature comfort ratio adjusted by the HVAC system. The calculation of $r^{bat}_{i, t}$, $r^{hvac}_{i, t}$, and $T^{comf}_{i, t}$ are shown in the following equation. 

\begin{equation}
\begin{aligned}
r^{bat}_{i, t} = \begin{cases} \frac{q^{bat}_{i, t}}{q^{bat}_{i, t} + q^{hvac}_{i, t}}, & \mbox{if battery charges} \\ 0, & \mbox{if battery discharges} \end{cases}
\end{aligned}
\label{eq:battery share}
\end{equation}

\begin{equation}
\begin{aligned}
r^{hvac}_{i, t} = \begin{cases} \frac{q^{hvac}_{i, t}}{q^{bat}_{i, t} + q^{hvac}_{i, t}}, & \mbox{if battery charges} \\ 1, & \mbox{if battery discharges} \end{cases}
\end{aligned}
\label{eq:hvac share}
\end{equation}
where $q^{bat}_{i, t}$ and $q^{hvac}_{i, t}$ represent the bidding quantity for the battery and HVAC system, respectively.

\begin{equation}
\begin{aligned}
T^{comf}_{i, t} = max(\epsilon, 1 - |\frac{min(|T^{set}_{i,t} - T^{in}_{i,t}|, T^{max}_{i,t})}{T^{max}_{i,t}}|)
\end{aligned}
\label{eq:temperature comfort ratio}
\end{equation}
where $T^{set}_{i,t}$ is desired indoor temperature defined by prosumer $i$. $T^{in}_{i,t}$ is the indoor temperature. $T^{max}_{i,t}$ is the maximum accepted temperature difference between desired indoor and actual room temperatures. $\epsilon$ is a positive constant value close to 0.

The valuation function is the prosumer's hidden (private) information that is not known by others. It also indicates the control of DERs.

\section{Market Markov Game Formulation}
The bidding decision generation, UDA market clearing, and coordination of energy management can be formulated as a finite Partially Observable Markov Game (POMG) \cite{jiayi2008review} with discrete time steps. This market game involves $n$ market participants. It also defines a set of global states $S$, observations $O_{1:n}$, actions $A_{1:n}$ and reward function $R_{1:n}$ from each prosumer, and a state transition function $T_{tran}$. Specifically, the market auctioneer publishes the global states $S$ shared by all prosumers. The observations $O_i$ consist of public observations from the global states and private observations only known by prosumer $i$. The state transition function $T_{tran}$ defines the probability distribution from the current state to the next possible states. The time interval between two consecutive states is one auction period. 

The objective of each prosumer in this market is to learn a strategy to maximize its cumulative expected reward. Specifically, at auction round $t$, prosumer $i$ takes actions based on its observations $O_{i,t}$. The market then moves into the next state $S_{t+1}$, according to the state transition function $T_{tran}$ conditioned on the actions of all prosumers. Each prosumer calculates its reward $R_{i,t}$ based on $S_{t+1}$ and obtains new observations $O_{i,t+1}$ for the next auction around. The significant elements of the formulated market game are explained as follows.

\subsection{Observations}
The prosumer $i$ at auction round $t$ has its observations $O_{i,t} = [O_{i,t-1}^{pub}, O_{i,t}^{pri}, p_{sup, t}]$. The public observations $O_{i,t-1}^{pub}$ comprises the market clearing information $O_{i,t-1}^{pub, clear}$ and market statistic information $O_{i,t-1}^{pub, stat}$ of auction round $t-1$. Specifically, the market clearing information $O_{i,t-1}^{pub, clear}$ includes market clearing price $\lambda_{t-1}^p$ and market clearing quantity $\lambda_{t-1}^q$. The market statistic information $O_{i,t-1}^{pub, stat}$ includes seller and buyer ratio ($r_{t-1}^s$, $r_{t-1}^b$), total seller and buyer quantities ($q_{tot, t-1}^s$, $q_{tot, t-1}^b$), mean bidding prices of sellers and buyers ($m_{t-1}^s$,  $m_{t-1}^b$), the standard deviation of sellers' bidding prices and buyers' bidding prices ($std_{t-1}^s$, $std_{t-1}^b$). The private observations $O_{i,t}^{pri}$ consists its available energy capacity $q_{i,t}^{ava}$ (i.e., $q_{i,t}^{ava} = E_{i,t}^{bat} + E_{i,t}^{pv}$), and total load consumption prediction $q_{i,t}^{load}$ (i.e., $q_{i,t}^{load} = E^{hvac}_{i,t} + E^{base}_{i,t}$). Additionally, the electricity supplier's price $p_{sup, t}$ at auction round $t$ is also known by each prosumer.

\subsection{Actions}
\begin{figure}[tb]
\centering
\includegraphics[width=3.4in]{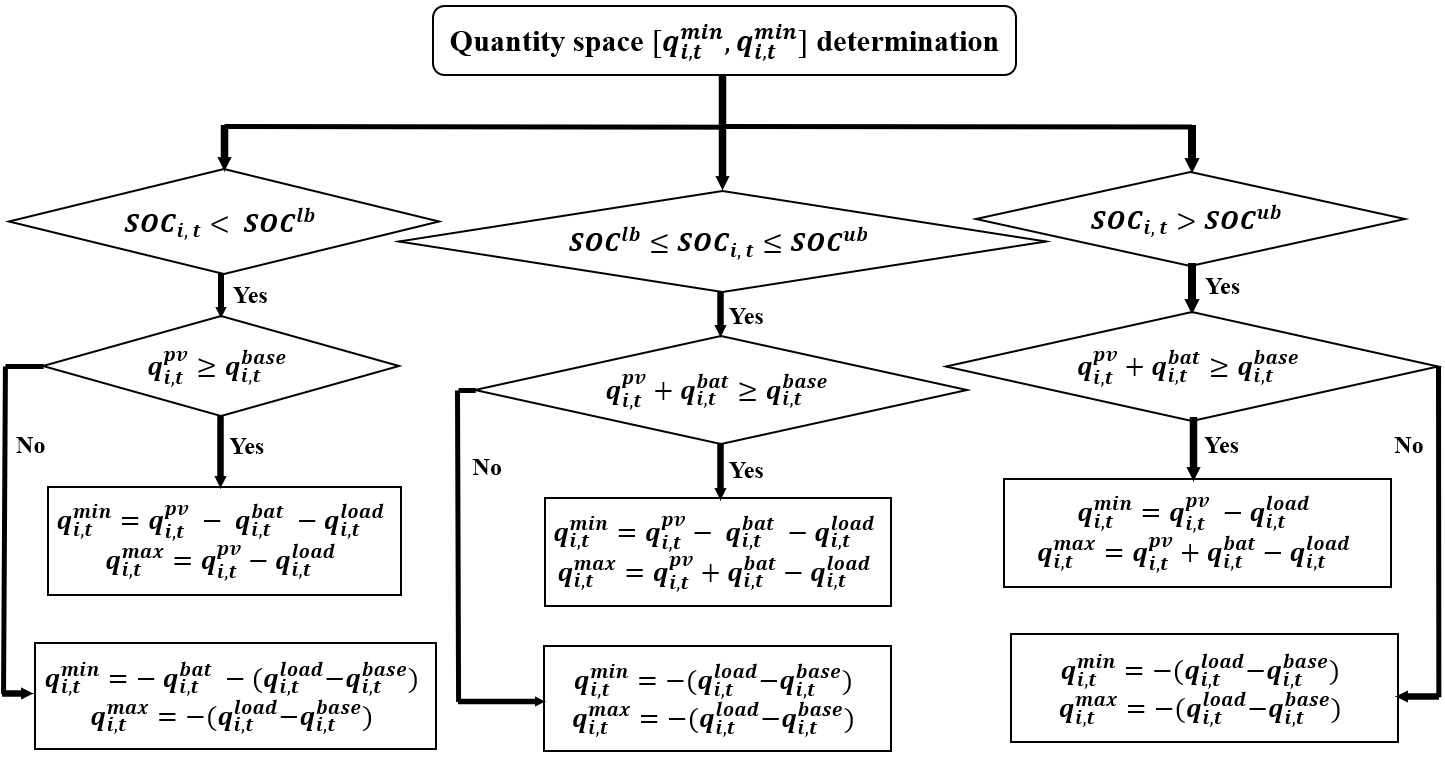}
\caption{Quantity space determination}
\label{fig:quantity_determination}
\vspace{-4mm}
\end{figure}

Prosumer $i$ at auction round $t$ generates its actions $A_{i, t} = [(p^b_{i,t} / p^s_{i,t}), q_{i, t}]$. The $p^b_{i, t} \in [0, 1] \mbox{ and } p^s_{i, t} \in [0, 1]$ represent the magnitude of buying and selling price submitted to the UDA market. The $q_{i, t}$ determines the magnitude of the selling quantity (positive), buying quantity (negative), or non-participant (0). Specifically, the space of bidding quantity $q_{i, t} \in [q_{i, t}^{min}, q_{i, t}^{max}] \subseteq [-10, 10]$ is dynamically changed based on the energy generation, storage, and consumption of the building. The determination of quantity space of prosumer $i$ at auction round $t$ is shown in Fig. \ref{fig:quantity_determination}. Furthermore, after the energy is dispatched by the market, the operation of each DER is subject to the determination process of the bidding quantity. Please note that, in this work, we set $p^b_{i, t}$ and $p^s_{i, t}$ as continuous variables and $q_{i, t}$ as a discrete variable, which is different from existing literature. 

\subsection{State Transition}
In this market game, the transition is not only affected by the actions of all prosumers but also by the inherent randomness and uncertainty of the system. The exogenous features composed of system intrinsic uncertainties (e.g., supplier's pricing, weather, etc.) are not changed by the prosumer's actions. Using a probabilistic model to accurately represent the transition presents a significant challenge. RL provides model-free solutions that can deal with these potential uncertainties. On the other hand, the state transition for the endogenous features is determined by the prosumers' actions at each auction round. The market clearing information (i.e., $[\lambda_{t}^p, \lambda_{t}^q]$) and market statistic information (i.e., $[r_{t}^s, r_{t}^b, q_{tot, t}^s, q_{tot, t}^b, m_{t}^s, m_{t}^b, std_{t}^s, std_{t}^b]$) are calculated based on UDA market clearing mechanism after all bids are received from market participants. Additionally, at auction round $t$, the energy in battery $E^{bat}_{i, t+1}$ for prosumer $i$ is determined by $C^{bat}_{i, t}$ and $D^{bat}_{i, t}$ after market cleared according to Equation (\ref{eq:battery_opt}). $C^{bat}_{i, t}$ and $D^{bat}_{i, t}$ are mutually exclusive since the charging and discharging of the battery cannot occur simultaneously at one auction round. 
\begin{equation}
\begin{aligned}
C^{bat}_{i, t} = q^d_{i,t} - q^{base}_{i, t} - q^{hvac}_{i, t} + q^{pv}_{i,t}
\end{aligned}
\label{eq:battery charge state}
\end{equation}
\begin{equation}
\begin{aligned}
D^{bat}_{i, t} = q^d_{i,t} + q^{base}_{i, t} + q^{hvac}_{i, t} - q^{pv}_{i,t}
\end{aligned}
\label{eq:battery discharge state}
\end{equation}
where the sign of dispatched quantity $q^d_{i,t}$ is determined by the prosumer's role at auction round $t$. Furthermore, based on the HVAC system model in \cite{du2017energy}, the indoor temperature $T^{in}_{i, t+1}$ for prosumer $i$ at auction round $t+1$ is determined by the indoor temperature and the outdoor temperature at step $t$, as well as the energy demand $q^{hvac}_{i, t}$ of the HVAC system

\subsection{Reward}
Each prosumer $i$ plays the role of a buyer, a seller, or a non-participant in each auction round. The prosumer $i$ also has different benefit considerations for different roles, leading to different reward strategy designs. 

When the prosumer $i$ acts as a seller at auction round $t$, it mainly considers the profit obtained from the TEM by selling energy. Its reward function can be calculated according to Equation (\ref{eq:reward seller}) after the UDA market is cleared. Specifically, when prosumer $i$'s selling bids are accepted by the market, its reward is the difference between utility gain (e.g., total revenue and base load payment savings) and utility loss (cost) due to its expected energy generation. Otherwise, the reward is simply regarded as a penalty for the production costs of wasted energy.
\begin{equation}
\begin{aligned}
R^s_{i,t} = \lambda_{t}^p q^d_{i,t} + \lambda_{t}^p q^{base}_{i, t} - (C_{pv} q_{i,t}^{pv} + C_{bat} q_{i,t}^{bat})
\end{aligned}
\label{eq:reward seller}
\end{equation}
where $q^d_{i,t}$, $q_{i,t}^{pv}$ and $q_{i,t}^{bat}$ represent the market dispatched quantity, predicted quantity generated by PVs and quantity from the battery, respectively. 

When prosumer $i$ plays a buyer role at auction round $t$, and its buying bid is accepted by the market, its reward can be the difference between its valuation value and expected payment as shown in Equation (\ref{eq:reward buyer}). Otherwise, its reward is calculated as the difference between the expected comfort sacrifice and payment saving. 
\begin{equation}
\begin{aligned}
R^b_{i,t}  = 
\begin{cases} 
 v_{i, t}(q_{i,t}^{d}) - \lambda^{p}_{t} q_{i,t}^{d}, \mbox{if } \lambda_{t}^p <= p^{b}_{i,t} \\
-v_{i, t}(q_{i,t}) + p^{b}_{i,t} q_{i,t}, \mbox{if } \lambda_{t}^p > p^{b}_{i,t} 
\end{cases}
\end{aligned}
\label{eq:reward buyer}
\end{equation}
where $p^{b}_{i,t}$, $v_{i, t}(q_{i,t}^{d})$, $v_{i, t}(q_{i,t})$ and $q_{i,t}^{d}$ represent the buying bid price, the valuation of a buyer based on the market dispatched quantity, the valuation of a buyer based on the bidding quantity and the market dispatched quantity, respectively.

When the prosumer $i$ is a non-participant player at auction round $t$, it uses energy from PVs and batteries to cover base and HVAC system loads. Therefore, its reward calculation involves the valuation value, base load payment savings and energy generation and operation cost. 
\begin{equation}
\begin{aligned}
R^n_{i,t} = v_{i, t}(q_{i,t}^{pv}+q_{i,t}^{bat}) + \lambda_{t}^p q^{base}_{i, t} - (C_{pv} q_{i,t}^{pv} + C_{bat} q_{i,t}^{bat})
\end{aligned}
\label{eq:reward np}
\end{equation}

In each episode containing $N$ auction rounds, the calculation of the overall reward of the prosumer $i$ is shown in Equation (\ref{eq:total reward}). Please note that, for each auction round, each prosumer can only play one role in participating in the market auction.
\begin{equation}
\begin{aligned}
R_{i} = \sum_{t=1}^{N} (R^s_{i,t} + R^b_{i,t} + R^n_{i,t}) 
\end{aligned}
\label{eq:total reward}
\end{equation}

\section{Proposed Reinforcement Learning Model}

Since the Deep Deterministic Policy Gradient (DDPG) algorithm can be applied to high-dimensional and continuous state and action spaces \cite{silver2014deterministic, lillicrap2015continuous}, it can effectively assist the agent in handling the designed observations and optimizing bidding strategies. Furthermore, because the training and execution of the concurrent learning framework are distributed (i.e., done by each agent in a private setting), it can protect user privacy and has good scalability. Therefore, the RL framework is based on the concurrent DDPG model in this work.

\subsection{Challenges}\label{sec:challenge} 
Some unique characteristics of our system require additional customized design of the DDPG model. \textit{First}, unlike most existing works that focus only on either optimal bidding prices or optimal quantity strategies, this work aims to achieve the optimal bidding solution by considering both factors. The hypothesis is that by adjusting both factors in a coordinated way, the resulting solution can facilitate prosumers to achieve higher utility. Specifically, this work adopts PE technology for bidding quantity design and DER control due to its advantages in improving the flexibility and cost-effectiveness of DERs. As a result, the bidding quantity in this work is set as a discrete value. However, since the bidding prices should be continuous values to provide prosumers more flexibility, we need to handle the challenge of a hybrid action space with continuous prices and discrete quantities. \textit{Second}, since a large number of prosumers learn their policies independently, such frequent policy changes can easily cause instability in the environment and make it very difficult to converge. 

To address these challenges, we propose a novel concurrent deep reinforcement learning framework with a set of shared, non-sensitive, learnable information among prosumers. In the following section, we introduce the proposed DRL model in detail.

\subsection{Concurrent Deep Reinforcement Learning Framework}
\begin{figure*}[tb]
\begin{center}
\includegraphics[width=0.9\textwidth,height=0.5\textwidth]{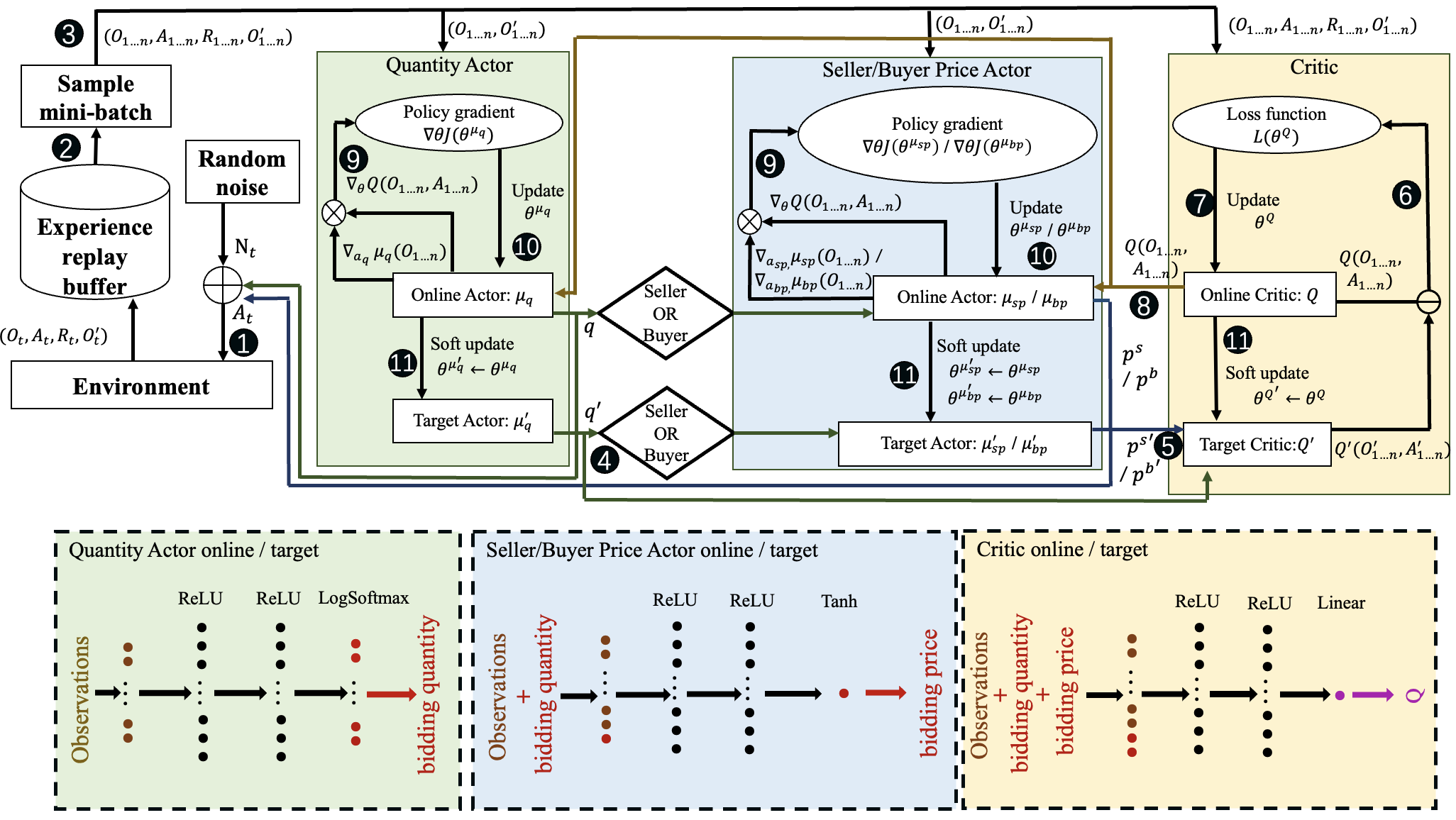}
\end{center}
\caption{The architecture of proposed concurrent deep reinforcement learning model}
\label{fig:ddpg}
\end{figure*}

In this study, each prosumer is modeled as a DDPG agent that integrates both public information shared among all prosumers and private information obtained by its own observations. Each agent contains two types of networks: actor networks and critic network, as shown in Fig. \ref{fig:ddpg}. The input and hidden layers of the actor and the critic networks are fully connected layers with the ReLU activation function. The actor networks map the observation of an agent to optimal actions. 

To address the hybrid action space challenge discussed above (in section \ref{sec:challenge}), we design two types of actor networks, which are quantity actor networks and price actor networks. The output of the quantity actor network indicates the bidding quantity and role by using the LogSoftmax activation function in the output layer. Furthermore, two types of roles (seller and buyer) need to offer bidding prices. However, each prosumer only generates energy from solar panels during sunny periods; the number of times a prosumer acts as a buyer in a day is significantly more than the number of times a prosumer acts as a seller. It causes a large difference in the content and the number of data samples for the two roles. Therefore, there are two types of price actor networks: the selling price actor network and the buying price actor network. The output of the quantity actor network determines which network from the seller/buyer price actor network is selected. When the output of the quantity actor is a non-participant, no price actor network is selected. The output of both price actor networks is a bidding price using the Tanh activation function in the output layer. The critic network evaluates actions from the actor networks to improve the performance of the actor networks. The output layer of the critic network uses a linear activation function.

In the training phase, the quantity actor network integrates the private local observations, public market clearing, and statistic information to calculate the optimal energy quantity to buy or sell. When the role of the output of the quantity actor network is buyer, the buying price actor network uses the local observation, the output of the quantity actor network, and market clearing and statistic information to calculate the optimal buying price. Otherwise, the selling price actor network uses local observation, the output of the quantity actor network, and market clearing and statistical information to calculate the optimal selling price. Furthermore, the critic network uses private local observations, actions from actor networks, public market clearing and statistical information to evaluate the actions. Please note that the market statistic information contains the actions of all agents as well as other non-sensitive statistics of the agents in the environment; its use by the critic network can play an essential role in keeping the environment stable, which addresses the convergence challenge discussed above (in Section \ref{sec:challenge}).  

In the execution phase, the quantity actor network and price actor networks are used while the critic network is removed. In addition, the UDA market also participates cooperatively as part of the environment, whose work is to calculate market clearing and market statistics information based on the received agent actions. 

\subsection{Learning Algorithm}
The proposed concurrent deep reinforcement learning algorithm is illustrated in Fig. \ref{fig:ddpg} and Algorithm \ref{alg:DDPG Algorithm}. 

Let the online quantity actor network, online selling price actor network, online buying price actor network, and online critic network of the agent be denoted as $\mu_{q}$, $\mu_{sp}$, $\mu_{bp}$ and $Q$ respectively. The weights of these networks are $\theta^{\mu_{q}}$,  $\theta^{\mu_{sp}}$, $\theta^{\mu_{bp}}$ and $\theta^{Q}$, respectively. Before training starts, the online actor and critic networks are created, and their weights are randomly initialized. The network topology of the target quantity actor network $\mu_{q}^{'}$, target selling price actor network $\mu_{sp}^{'}$, target buying price actor network $\mu_{bp}^{'}$ and target critic network $Q^{'}$ of the agent are the same as the topology of the corresponding online actor networks ($\mu_{q}$, $\mu_{sp}$, $\mu_{bp}$) and critic network ($Q$). The network weights of target actor networks and target critic network are initialized as $\theta^{\mu_{q}^{'}}$ $\leftarrow$ $\theta^{\mu_{q}}$, $\theta^{\mu_{sp}^{'}}$ $\leftarrow$ $\theta^{\mu_{sp}}$, $\theta^{\mu_{bp}^{'}}$ $\leftarrow$ $\theta^{\mu_{bp}}$, and $\theta^{Q^{'}}$ $\leftarrow$ $\theta^{Q}$. For each agent, a replay buffer $Buf$ is created and initialized to store list of tuples ($O$, $A$, $R$, $O^{'}$).

For each training episode, a random action exploration process and an observation are initialized. After receiving observations $O_t$ and noise $N_t$, the agent takes actions $A_t = [q_{t}, (p^s_{t} / p^b_{t})]$ according to Equation (\ref{eq:quantity_action}) and (\ref{eq:price_action}) at step $t$ (process 1 of Fig. \ref{fig:ddpg}). A non-participant agent's actions can be represented as $A_t = [0, 0]$.

\begin{equation}
\begin{aligned}
q_t = \mu_q(O_t) + N_t,
\end{aligned}
\label{eq:quantity_action}
\end{equation}
\begin{equation}
\begin{aligned}
p^{s/b}_t = \mu_{sp/bp}(O_t, q_t) + N_t,
\end{aligned}
\label{eq:price_action}
\end{equation}

At each step $t$, the actions generated by actor networks of each agent are submitted to the TEM, where the UDA market clearing mechanism calculates market clearing and statistic information. The UDA market then feeds this information back to all agents. After receiving all the market information, each agent calculates its reward $R_t$ based on its role according to Equation (\ref{eq:reward seller}, \ref{eq:reward buyer}, \ref{eq:reward np}), and updates the next observation $O_{t}^{'}$. The tuple ($O_t$, $A_t$, $R_t$, $O^{'}_{t}$) is stored in replay buffer $Buf$ and observation $O_t$ is updated by $O_{t}^{'}$.

After every fixed number of steps, each agent randomly samples $n$ number of transitions ($O_{1,..,n}$, $A_{1,..,n}$, $R_{1,..,n}$, $O^{'}_{1,...,n}$) from the replay buffer (process 2 and 3 of Fig. \ref{fig:ddpg}) to train the actor networks and critic network by updating the network weights of the online and target networks. 

Each agent calculates the loss of the online critic network $\theta^{Q}$ by Equation (\ref{eq:critic_loss}) (process 4 to 6 in Fig. \ref{fig:ddpg}). Then, it updates the weights of the online critic network $\theta^{Q}$ by Equation (\ref{eq:critic_target_update}) (process 7 in Fig. \ref{fig:ddpg}).
\begin{equation}
\begin{aligned}
L(\theta^Q) = \frac{1}{n} \sum_{i=1}^n (y_i - Q(O_i, A_i | \theta^Q))^2 \\
y_i = R_i + \gamma Q^{'}(O^{'}_i, A^{'}_i | \theta^{Q^{'}})
\end{aligned}
\label{eq:critic_loss}
\end{equation}
\begin{equation}
\begin{aligned}
\theta^{Q} \leftarrow \theta^{Q} + \alpha^{Q} \nabla_{\theta^{Q}} L(\theta^Q) \\
\end{aligned}
\label{eq:critic_target_update}
\end{equation}
where $Q(O_i, A_i | \theta^Q)$ and $Q^{'}(O^{'}_i, A^{'}_i | \theta^{Q^{'}})$ are the predicted Q-values from online critic network and target critic network, respectively. The actions $A_i$ and $A^{'}_i$ are the predicted actions from online actor networks and target actor networks, respectively. $y_i$ is the target Q-value and $R_i$ is the reward. $\gamma$ is the discount factor. $\alpha^{Q}$ is the learning rate of the gradient descent algorithm.

Each agent calculates its sampled policy gradients for online actor networks according to Equation (\ref{eq:actor_update}) (process 8 and 9 in Fig. \ref{fig:ddpg}). 
\begin{equation}
\begin{aligned}
\nabla_{\theta^{\mu}} J(\theta^{\mu}) = \frac{1}{n} \sum_{i=1}^n \nabla_{\theta^{\mu}} \mu (O_i | \theta^{\mu})  \nabla_{\mu(O_i)} Q(O_i, A_i | \theta^Q)
\end{aligned}
\label{eq:actor_update}
\end{equation}
where $\mu$ in Equation (\ref{eq:actor_update}) can be represented as $\mu_{q}$, $\mu_{sp}$ and $\mu_{bp}$.

The following updates are then applied to the weights of $\mu_{q}$, $\mu_{sp}$, and $\mu_{bp}$ (process 10 in Fig. \ref{fig:ddpg}) respectively.
\begin{equation}
\begin{aligned}
\begin{cases}
\theta^{\mu_{q}} \leftarrow \theta^{\mu_{q}} + \alpha^{\mu_{q}} \nabla_{\theta^{\mu_q}} J(\theta^{\mu_q}) \\ 
\theta^{\mu_{sp}} \leftarrow \theta^{\mu_{sp}} + \alpha^{\mu_{sp}} \nabla_{\theta^{\mu_{sp}}} J(\theta^{\mu_{sp}}) \\ 
\theta^{\mu_{bp}} \leftarrow \theta^{\mu_{bp}} + \alpha^{\mu_{bp}} \nabla_{\theta^{\mu_{bp}}} J(\theta^{\mu_{bp}}) \\
\end{cases}
\end{aligned}
\label{eq:online actor networks update}
\end{equation}
where $\alpha^{\mu_{q}}$, $\alpha^{\mu_{sp}}$ and $\alpha^{\mu_{bp}}$ are the learning rates of the gradient decent algorithm.

Furthermore, the target actor networks ($\theta^{\mu_{q}^{'}}$, $\theta^{\mu_{sp}^{'}}$, $\theta^{\mu_{bp}^{'}}$) and critic network $\theta^{Q^{'}}$ are updated according to Equation (\ref{eq:target networks update}) (process 11 in Fig. \ref{fig:ddpg}).
\begin{equation}
\begin{aligned}
\begin{cases} 
\theta^{\mu_{q}^{'}} \leftarrow \tau\theta^{\mu_{q}} + (1 - \tau)\theta^{\mu_{q}^{'}} \\ 
\theta^{\mu_{sp}^{'}} \leftarrow \tau\theta^{\mu_{sp}} + (1 - \tau)\theta^{\mu_{sp}^{'}} \\ 
\theta^{\mu_{bp}^{'}} \leftarrow \tau\theta^{\mu_{bp}} + (1 - \tau)\theta^{\mu_{bp}^{'}} \\
\theta^{Q^{'}} \leftarrow \tau\theta^{Q} + (1 - \tau)\theta^{Q^{'}} \\
\end{cases}
\end{aligned}
\label{eq:target networks update}
\end{equation}
where $\tau$ is the soft update coefficient of the target network.

\begin{algorithm}[tb]
\caption{Proposed DRL model}
\label{alg:DDPG Algorithm}
\LinesNumbered 
Initialize online critic and actor networks. \\
Initialize target critic and actor networks. \\
Initialize the learning rate of critic and actor networks. \\
Initialize the update rate of critic and actor networks. \\
Initialize replay buffer and sample batch size. \\

\For{episode = 1 : N}{
Initialize a random process for action exploration. \\
Reset the environment and obtain initial observation $O$. \\

\For{t = 1 : T}{
\ForEach{$agent \in Agents$}{
Determine actions $A$. \\
Calculate the reward $R$. \\
Deposit experience ($O$, $A$, $R$, $O^{'}$).\\ 
Update $O$ $\leftarrow$ $O^{'}$. \\

\If{$mod(t, update\ rate) = 0$}
{
Randomly sample mini-batch. \\
Update the online critic network by minimizing $L(\theta^Q)$. \\
Update the online actor networks using policy gradient $\nabla_{\theta^{\mu_q}} J(\theta^{\mu_q})$, $\nabla_{\theta^{\mu_{sp}}} J(\theta^{\mu_{sp}})$ and $\nabla_{\theta^{\mu_{bp}}} J(\theta^{\mu_{bp}})$. \\
Update target critic and actor networks. \\
}

}
}
}
\end{algorithm}

\section{Experiment Setup and Results Analysis}
\subsection{Experiment Setup}
The designed TEM with DER-equipped prosumers are simulated by the PEMT-CoSim platform developed by our prior work \cite{li2022pemt}, which is a co-simulation platform for TEM to investigate the packetized energy in a smart distributed system based on Transaction Energy Simulation Platform (TESP) \cite{huang2018simulation}. The proposed DRL model is integrated into the PEMT-CoSim platform as an AI module. In the experiments, the UDA-based TEM consists of a market operator, 30 houses, and an energy supplier, which provides energy quantity with a varying wholesale market price ranging from 0.2 $\$/kWh$ to 0.4 $\$/kWh$. The parameters of the TEM and DRL model are summarized in TABLE \ref{tabel:Experiment settings of the proposed model}. 

\begin{table}[t]
\footnotesize
%\small
\caption{Experiment settings in the simulation}
\label{tabel:Experiment settings of the proposed model}
\begin{tabular}{|cc|}
\hline
\multicolumn{2}{|c|}{\textbf{Transactive energy market settings}} \\ \hline
\multicolumn{1}{|c|}{Simulation days} &
  60 \\ \hline
\multicolumn{1}{|c|}{Transactive energy market} &
  UDA-based \\ \hline
\multicolumn{1}{|c|}{Duration of auction} &
  300 seconds \\ \hline
\multicolumn{1}{|c|}{Number of residential buildings} &
  30 \\ \hline
\multicolumn{1}{|c|}{PV panel of each building} &
  \begin{tabular}[c]{@{}c@{}} Quantity: 6-14, \\ unit power: 480 W\end{tabular} \\ \hline
\multicolumn{1}{|c|}{Battery of each building} &
  \begin{tabular}[c]{@{}c@{}}charge/discharge rate: 3 kW, \\ capacity: 10kWh, \\ SOC range: 0.1-0.8\end{tabular} \\ \hline
\multicolumn{1}{|c|}{HVAC system power} &
  3 kW \\ \hline
\multicolumn{2}{|c|}{\textbf{Deep reinforcement learning model settings}} \\ \hline
\multicolumn{1}{|c|}{\begin{tabular}[c]{@{}c@{}}Learning rate of critic and \\ actor neural networks\end{tabular}} &
  0.0001 \\ \hline
\multicolumn{1}{|c|}{\begin{tabular}[c]{@{}c@{}}Discount factor of \\ critic neural network\end{tabular}} &
  0.9 \\ \hline
\multicolumn{1}{|c|}{Exploration noise} &
  \begin{tabular}[c]{@{}c@{}}Gaussian \\ (with uadratic noise decay)\end{tabular} \\ \hline
\multicolumn{1}{|c|}{\begin{tabular}[c]{@{}c@{}}Soft update coefficient of \\ target neural networks\end{tabular}} &
  0.001 \\ \hline
\multicolumn{1}{|c|}{Minibatch size} &
  64 \\ \hline
\multicolumn{1}{|c|}{Training / Testing data} &
  54 days / 6 days \\ \hline
\multicolumn{1}{|c|}{Optimizer} &
  Adam \\ \hline
\multicolumn{1}{|c|}{\begin{tabular}[c]{@{}c@{}}Number of neurons \\ (hidden layer)\end{tabular}} &
  200 \\ \hline
\multicolumn{1}{|c|}{\begin{tabular}[c]{@{}c@{}}Activation function \\ (hidden layer)\end{tabular}} &
  ReLU \\ \hline
\multicolumn{1}{|c|}{\begin{tabular}[c]{@{}c@{}}Activation function  \\ (output layer, critic network)\end{tabular}} &
  Linear \\ \hline
\multicolumn{1}{|c|}{\begin{tabular}[c]{@{}c@{}}Activation function \\ (output layer, price actor networks) \end{tabular}} &
  Tanh \\ \hline
\multicolumn{1}{|c|}{\begin{tabular}[c]{@{}c@{}}Activation function o\\ (output layer, quantity actor  network) \end{tabular}} &
  LogSoftmax \\ \hline
\multicolumn{1}{|c|}{Deep learning framework} &
  PyTorch V1.13.1 \\ \hline
\end{tabular}
\end{table}

\subsection{Results Analysis}
This section evaluates the performance of the proposed model and then compares it with other state-of-the-art models. 

\subsubsection{Performance of proposed model}
In this subsection, the performance of the proposed model is evaluated from three perspectives: (1) selection of hyperparameters, (2) evaluation of comfort satisfaction and energy consumption of prosumers, (3) analysis of the impact of prosumer's role and bidding strategies on TEM. All experiments were applied with the same experimental settings.

\begin{figure}[tb]
\centering
\includegraphics[width=3.5in]{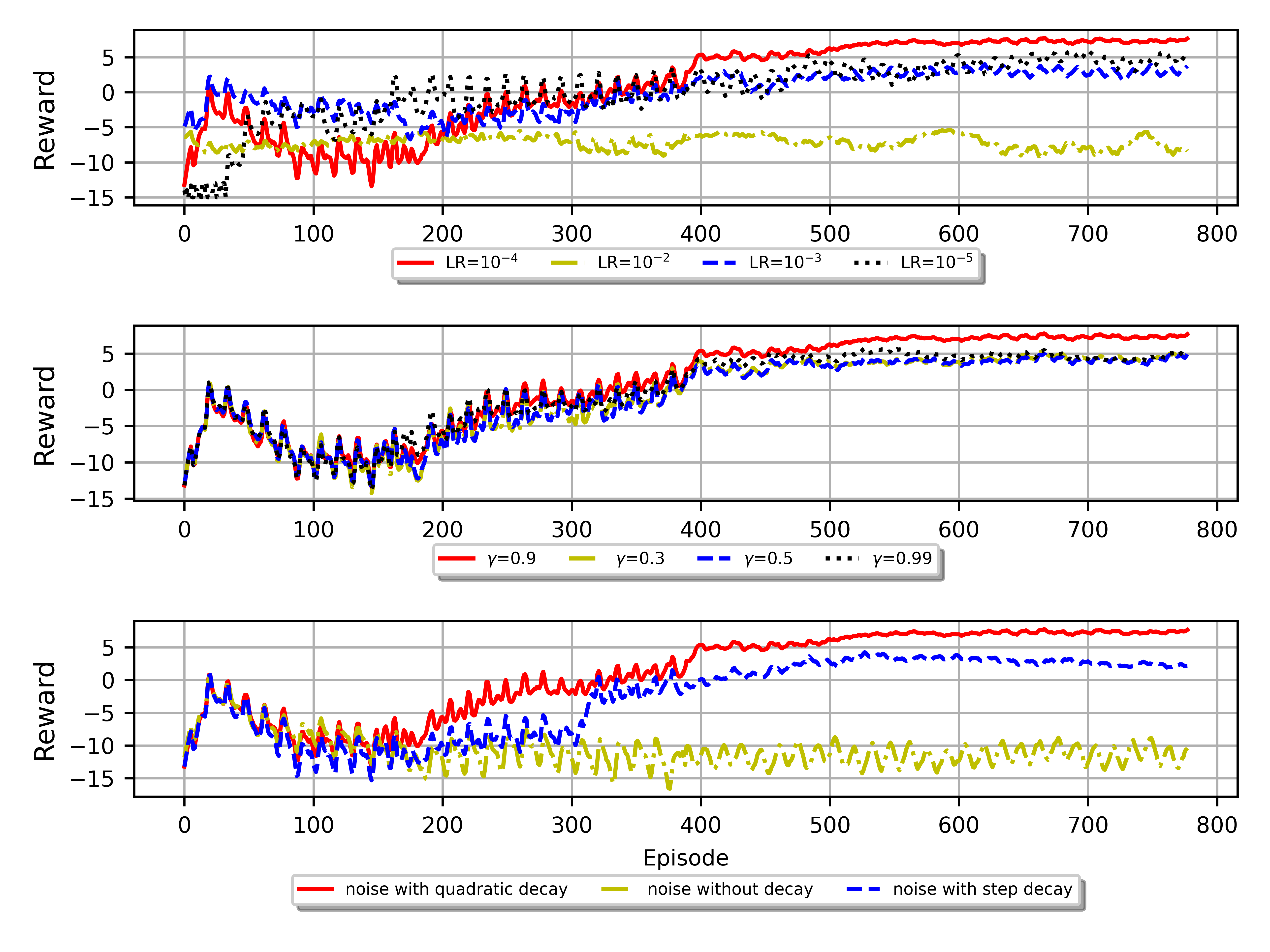}
\caption{Episode reward of the proposed model with different learning parameters}
\label{fig:module_test}
\vspace{-4mm}
\end{figure}

\paragraph{Selection of hyperparameters} 
The lines in Fig. \ref{fig:module_test} represent the average episode reward of thirty prosumers in the market. 

The first sub-figure of Fig. \ref{fig:module_test} shows the episode reward of the proposed model with different learning rates. When the learning rate of the neural network is large (yellow dash-dotted line), the network may fail to converge. On the other hand, when the learning rate of the neural network is small (black dashed line), the network may converge only to a local optimum. Specifically, in our experiment scenarios, the ideal learning rates of the actor networks and critic network are both 0.0001 (red solid line), which not only achieves the convergence of reward but also reaches the largest reward value.

The second sub-figure of Fig. \ref{fig:module_test} shows the episode reward of the proposed model with different values of $\gamma$, $\gamma$ is used as a discount factor to measure the importance of future rewards. When $\gamma$ approaches 1, the agent considers future rewards more than immediate ones. Prosumers not only calculate the immediate reward obtained by taking the bidding actions but also consider the impact of these actions on future rewards. When the value of $\gamma$ is too low (e.g., $\gamma = 0.3$, yellow dash-dotted line), the proposed model becomes unsightly. When the value of $\gamma$ is too high (e.g., $\gamma = 0.99$, black dotted line), the proposed model over-considers future rewards and may produce poor quality policies due to the divergent Q-value function estimation. The discount factor $\gamma$ with 0.9 (red solid line) shows the best performance of episode rewards. 

In the first sub-figure of Fig. \ref{fig:module_test}, the episode reward based on different noise strategies for action selection is compared. The Gaussian noise $N(0,\sigma^{2})$ is used to help the agent explore the optimal bidding behavior in these experiments. In the early stages of learning, since the agent has less knowledge about the environment, the noise needs to be set large to encourage the agent to explore the action space. However, as the learning period increases, the noise should be gradually reduced because the agent can use the accumulated experience to generate rewarding actions. Specifically, the decay of exploration noise is applied on $\sigma$. Noise without decay (yellow dash-dotted line) demonstrates that the introduction of large noises at all stages of learning leads to the agent being unable to use the experience to determine an optimal bidding strategy. On the other hand, noise with step decay (blue dashed line) drops the noise by reducing $\sigma$ every 100 episodes. Noise with time-based decay (red solid line) updates the noise by decreasing $\sigma$ in each step. Noise with step decay and time-based decay strategies help the agent obtain higher rewards. The time-based quadratic decay performs better because of its smooth linear decrease.

\begin{figure}[tb]
\centering
\includegraphics[width=3.5in]{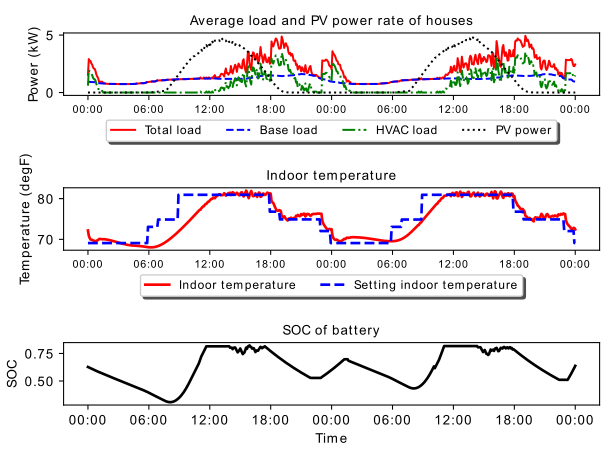}
\caption{Example of average load, PV power, indoor temperature, and SOC of the battery of houses for two days}
\label{fig:house_temp_information}
\vspace{-4mm}
\end{figure}

\paragraph{Evaluation of comfort satisfaction and energy operation of prosumers}
Fig. \ref{fig:house_temp_information} shows the average load consumption and PV power rate, indoor temperature changing, and SOC of the battery of all houses over two days. In the sub-figure (average load and PV power rate of houses), the total load increases due to the increased HVAC load caused by the increase in temperature. Additionally, the mean PV power rate (black dotted line) increases with increasing sunlight intensity during the day. In the sub-figure (indoor temperature), the red solid line and blue dashed line represent the actual indoor temperature and setting indoor temperature of houses. Specifically, the indoor temperature matches the setting temperature well. It means that the bidding demands are successfully dispatched by the TEM so that the HVAC system works as required. It validates that our proposed model can effectively meet the requirement of comfort satisfaction. In the sub-figure (SOC of houses), the SOC of the battery changes with the change of market participation roles caused by PV power rate, total load, and market electricity price.

%In the average SOC of houses subplot, as the PV power rate increases, the average SOC of the battery increases. The difference between the maximum SOC and the minimum SOC is because the houses choose different battery operations, which are based on the optimal bidding quantity strategy from the proposed model (quantity actor). For example, some houses choose to charge the battery to reduce the cost of future purchases instead of selling the surplus energy to obtain immediate rewards. Please note that the house's battery operations are a reward-maximizing choice based on its condition.

%\begin{figure}[tb]
%\centering
%\includegraphics[width=3.5in, trim={0 45 0 0}, clip]{house_load_information.png}
%\caption{Example of average load, PV power, indoor temperature and SOC of battery of houses for two days}
%\label{fig:house_temp_information}
%\vspace{-4mm}
%\end{figure}

\paragraph{Analysis of the impact of prosumer's role and bidding strategies on TEM}

\begin{figure}[tb]
\centering
\includegraphics[width=3.5in]{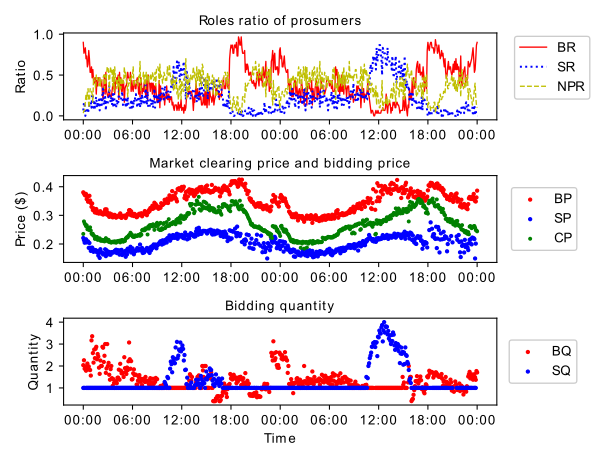}
\caption{Example of average prices (bidding prices, wholesale price, and market clearing price) and roles ratio of prosumers in TEM for two days}
\label{fig:market price change}
\vspace{-4mm}
\end{figure}

Fig. \ref{fig:market price change} shows bidding prices, bidding quantity, and roles ratio of market participants over two days. As shown in the roles ratio of market participants sub-figure, during the day, as the PV power rate increases, the seller ratio (SR, blue dotted line) increases, and prosumers' batteries are charged. When the house has enough available energy (e.g., energy in the battery, energy generated by PV) to cover the load consumption, it tends to become a non-participant (NPR, yellow dashed line) to avoid energy payment costs and reduce transmission pressure on the grid. Otherwise, the house participates in the TEM as a buyer (BR, red solid line) to ensure the energy consumption of its appliances (e.g., HVAC system). 

In the price sub-figure, the red, blue, and green dots represent the average buying bidding price (BP), average selling bidding price (SP), and market clearing price (CP) at each auction round. The change in the market clearing price in the figure is determined by the amount of energy purchased or sold by buyers and sellers in TEM and the corresponding price according to the UDA market mechanism. 

In the bidding quantity sub-figure, the red and blue dots represent the average buying bidding quantity (BQ) and average selling bidding quantity (SP) at each auction round. Specifically, selling bidding quantity increases with the increase of PV power rate. During periods of higher electricity prices (e.g., 18:00 to 21:30), some houses join the market as buyers, primarily due to an increase in HVAC load. However, these houses purchase relatively less energy from the market because the proposed model reduces energy purchases by discharging the battery. When houses participate as buyers in the market during periods of lower energy prices, the proposed model guides them to increase their purchasing quantity.

\subsubsection{Overall performance comparison of proposed model and existing models}
This subsection compares the proposed model with a baseline model and two RL models adopted in the existing literature. Specifically, in the baseline model, the prosumer randomly selects the energy quantity to buy or sell based on the energy range allowed by the house and randomly selects the bidding price within a reasonable price range. In the other two RL-based models from \cite{namalomba2022agent, taghizadeh2022deep}, the bidding price as strategic action is generated by the Q-learning based model (Q-learning) and DDPG-based model (DDPG (price)), respectively. The buying or selling bidding quantity for each prosumer is randomly selected from the accepted demand range and available energy range provided based on house conditions. In addition, we add an experiment using the DDPG model (DDPG (quantity)) to determine the optimal bidding quantity strategy. In this experiment, the bidding price is set to a competitive price that guarantees that the bidding quantity of prosumers can be successfully dispatched by TEM. All comparison models are applied to the same market, prosumers, and weather environment settings.

\begin{figure}[tb]
\centering
\includegraphics[width=3.5in]{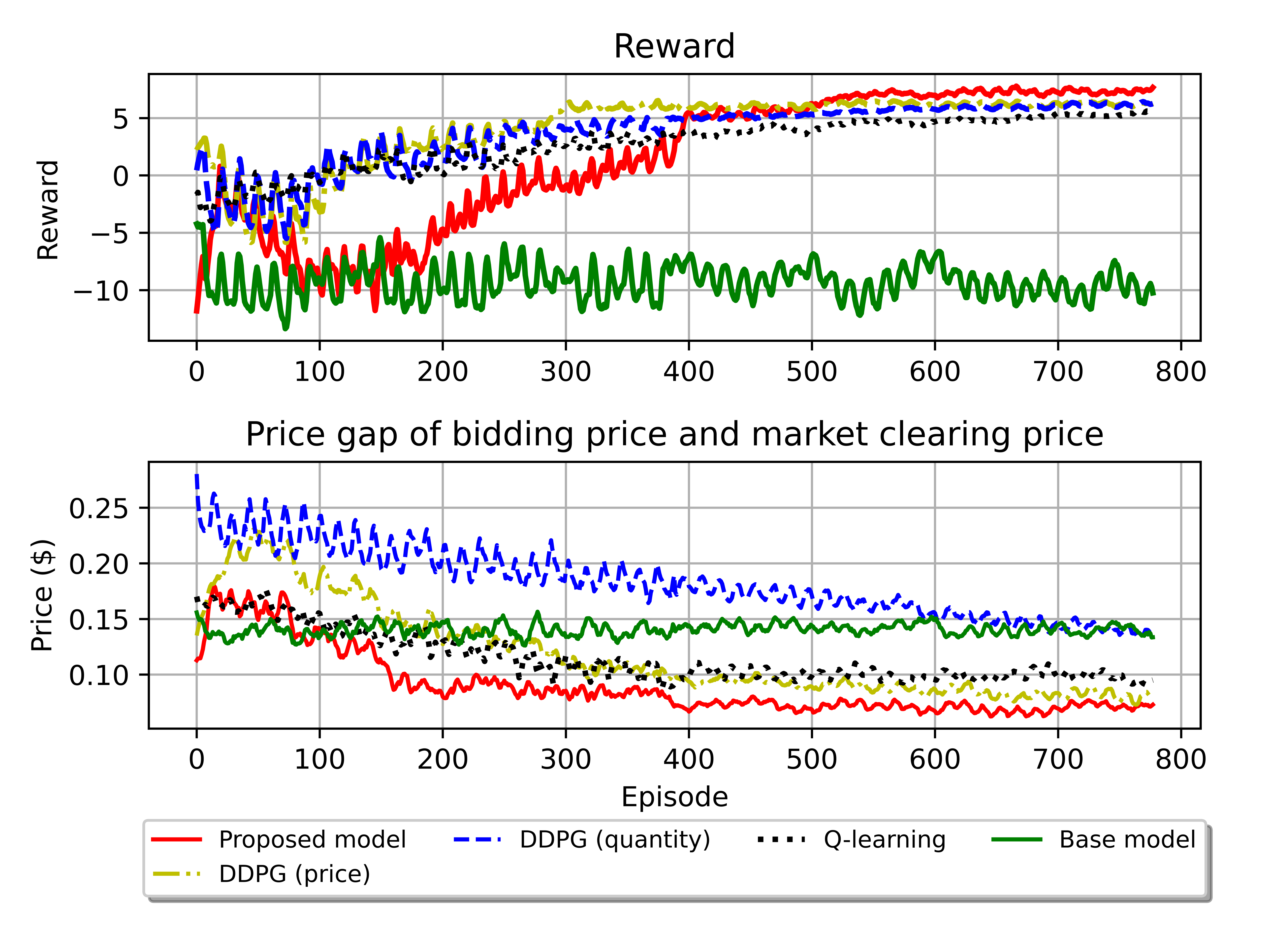}
\caption{Episode rewards and price gaps of baseline, DDPG (price), DDPG (quantity), Q-learning and proposed model}
\label{fig:models_rewards}
\vspace{-4mm}
\end{figure}

Fig. \ref{fig:models_rewards} shows the episode reward and the price gap between the bidding and market clearing prices. From the sub-figure of reward, it is observed that the proposed model achieves the highest rewards. The randomness strategy of the baseline model makes prosumers unable to obtain the required energy from the market, resulting in the loss of reward. Q-learning's insufficient ability to handle multi-dimensional state space and multi-iterative double auction market also results in fewer rewards for prosumers. Although DDPG (price) and DDPG (quantity) overcome the limitations of Q-learning by using neural networks, single types of bidding strategies still potentially result in the loss of maximum rewards. Compared with other models, the reward increase of the proposed model in the early stage of training is slower because the proposed model needs to train more neural networks than other models.

The price gap between the bidding and market clearing prices evaluates the clearing price prediction capability. Generally, the agent minimizes the gap between its bidding price and market clearing price. From the sub-figure of the price gap, it can be observed that the proposed model achieves the smallest price gap. The second best-performing model is the DDPG model (price). This is because these two models can output continuous price variables, predicting market prices more accurately. 

\begin{figure}[tb]
\centering
\includegraphics[width=3.5in]{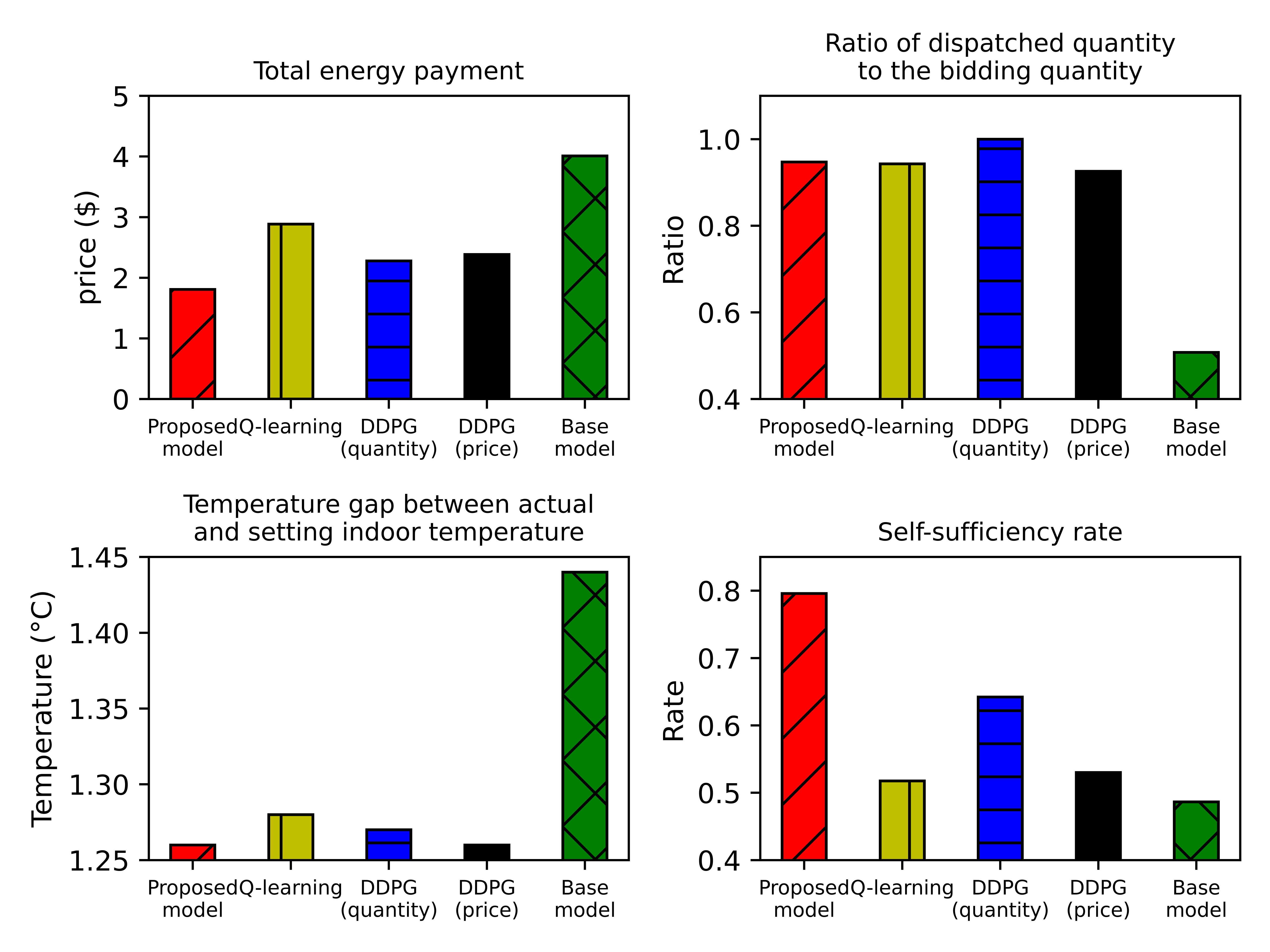}
\caption{Overall performance comparison of baseline, DDPG (price), DDPG (quantity), Q-learning and proposed model}
\label{fig:overall performance comparison}
\vspace{-4mm}
\end{figure}

Fig. \ref{fig:overall performance comparison} compares the daily performance of the proposed model and the existing models from four perspectives after training. Total energy payment is the cost of the consumer purchasing energy. The total energy payment of the proposed model is lower than that of Q-learning, DDPG (quantity), DDPG (price), and baseline models by 60\%, 26\%, 32\%, and 122\%. The temperature gap between actual and set indoor temperature evaluates the temperature comfort satisfaction. Except for the baseline model, the learning models can satisfy the prosumer's temperature comfort setting. The ratio of the dispatched quantity to the bidding quantity evaluates the ratio of successful bidding quantity. The ratio of the DDPG (quantity) model is 1.0 because of its competitive bidding price in the market, that is, high bidding buying price and low bidding selling price. It can be observed that the ratio of successful bidding quantity of the proposed model performs well. It validates that the proposed model with the seller/buyer price actor network can generate proper bidding prices acceptable to the market, enabling successful purchases or sales of energy.

The self-sufficiency rate evaluates the prosumer's independence in the external electricity market by calculating the ratio of its market participant role as seller or non-participant. A self-sufficiency rate can not only help the prosumer reduce the energy payment but also reduce the transmission pressure of the grid. The proposed model and DDPG (quantity) show a higher self-sufficiency rate. This is because the quantity actor networks in both models effectively assist prosumers in determining their roles and the corresponding energy quantities. Specifically, the self-sufficiency rate of the proposed model is higher than that of Q-learning, DDPG (quantity), DDPG (price), and baseline models by 54\%, 24\%, 50\%, and 64\%. 

\section{Conclusion}
In the smart grid, an effective DRL model in double auction-based TEM plays an important role in maximizing prosumers' economic profits and social welfare and reducing the stress on the distributed system. This paper first designs a UDA-based TEM, which consists of market operators and multiple and diverse DER-equipped prosumers as buyers or sellers. Furthermore, this paper proposes a DRL model based on distributed learning and execution to help prosumers make optimal energy trading decisions. Comprehensive experiments validate the effectiveness and robustness of the designed TEM and proposed model. The results show that, compared with other RL models adopted in the existing literature, the proposed model can help the prosumer make better bidding strategies. 

Future work will further improve the practicality of the proposed model (e.g., introducing more distribution system network constraints, such as voltage and current thermal limit). In addition, more prosumers and more complex DER operations (e.g., the optimal control strategy for each DER) will also be handled by the proposed model in future work.

\bibliographystyle{IEEEtran}
\bibliography{reference.bib}
\end{document}